\documentclass[prl,twocolumn,superscriptaddress,amsmath,amssymb]{revtex4-2}
\usepackage[colorlinks=true,citecolor=blue,linkcolor=blue,breaklinks=true]{hyperref}

\usepackage{color,graphicx,soul}
\usepackage{bm}
\usepackage{amsmath}
\usepackage{amssymb}
\usepackage{caption}
\usepackage[usenames,dvipsnames]{xcolor}
\usepackage{verbatim}
\usepackage{float}
\usepackage[justification=Justified]{caption}

\begin{document}
		
		\title{Andreev spin qubits bound to Josephson vortices in spin-orbit \\coupled planar Josephson junctions}

		\author{Katharina Laubscher}
		\affiliation{Condensed Matter Theory Center and Joint Quantum Institute, Department of Physics, University of Maryland, College Park, MD 20742, USA}
		
		\author{Valla Fatemi}
		\affiliation{School of Applied and Engineering Physics, Cornell University, Ithaca, NY, 14853, USA}
		
		\author{Jay D. Sau}
		\affiliation{Condensed Matter Theory Center and Joint Quantum Institute, Department of Physics, University of Maryland, College Park, MD 20742, USA}
		
		\date{\today}
		
		\begin{abstract}
	     We propose a variant of Andreev spin qubits (ASQs) defined in planar Josephson junctions based on spin-orbit coupled two-dimensional electron gases (2DEGs) in a weak out-of-plane magnetic field. The magnetic field induces a linear phase gradient across the junction, generating Josephson vortices that can host low-energy Andreev bound states (ABSs). We show that, in certain parameter regimes, the combined effect of the phase gradient and spin-orbit coupling stabilizes an odd-fermion parity ground state, where a single Josephson vortex binds a spinful low-energy degree of freedom that is energetically separated from the other ABSs. This low-energy degree of freedom can be exploited to define a special type of ASQ, which we dub the \emph{vortex spin qubit} (VSQ). We show that single-qubit gates for VSQs can be performed via flux driving, while readout can be achieved by adapting standard circuit quantum electrodynamics (cQED) techniques developed for conventional ASQs. We further outline how an entangling two-qubit gate can be performed using an ac current drive. We argue that VSQs offer prospects for a substantial reduction in device complexity and hardware overhead compared to conventional ASQ implementations, while preserving key advantages such as supercurrent-based readout, single-qubit gates, and long-range two-qubit gates.
		\end{abstract}
		
		\maketitle

\textit{Introduction.---}%
Andreev spin qubits (ASQs)~\cite{Chtchelkatchev2003,Padurariu2010} defined in short superconductor--normal--superconductor (SNS) Josephson junctions (JJs) have the potential to combine the complementary strengths of semiconductor spin qubits with those of superconducting qubits. The key idea is to store quantum information in the microscopic spin degree of freedom associated with a spin-split subgap Andreev bound state (ABS) in a JJ with strong spin-orbit coupling. In this way, ASQs retain the compact footprint of spin qubits, but at the same time benefit from a strong coupling to the macroscopic supercurrent through the junction. The latter allows for coherent manipulation of ASQs via a current or flux drive, while fast high-fidelity readout can be achieved using circuit quantum electrodynamics (cQED) techniques. Moreover, the supercurrent can mediate a fast, tunable interaction between distant ASQs, which is challenging to achieve in conventional spin qubit architectures. These advantages make ASQs a promising platform for scalable, high-speed, and strongly connected quantum hardware.

Conventional ASQs are defined in one-dimensional (1D) JJs based on proximitized semiconducting nanowires. In these structures, Andreev spin states can be stabilized either by (a) a charging energy~\cite{PitaVidal2023,Bargerbos2023,PitaVidal2024}, (b) a large external magnetic field~\cite{Whiticar2021}, or (c) non-equilibrium quasiparticle population~\cite{Hays2018,Hays2020,hays_coherent_2021}. While the first proof-of-principle realizations of ASQs~\cite{Hays2020,hays_coherent_2021} were based on option (c), this route is impractical because it inevitably results in a short qubit lifetime. Options (a) and (b) are in principle viable, but come with experimental challenges: On the one hand, to create a charging energy, it is necessary to at least partially reduce the tunnel coupling of the ASQ to the superconductors, which requires additional gate electrodes and control knobs for tune-up and has the potential to reduce the spin-dependent supercurrent. On the other hand, large magnetic fields create complications for superconducting devices~\cite{vanWoerkom2015,Song2009} by destabilizing the parent superconductor and decreasing the quasiparticle poisoning time, and reduce the usable volume in a cryogenic system.

To overcome these difficulties, it has been suggested that a \textit{three-terminal} device can stabilize an Andreev spin with only magnetic flux control~\cite{Heck2014,Coraiola2024,Svetogorov2025}, which requires only weak locally controllable magnetic fields routinely used in the best-performing conventional superconducting qubit systems~\cite{Google2025}. In this architecture, the phase winding around the three terminals strongly breaks time-reversal symmetry so that a spin state can be stabilized. However, this comes at the cost of a more complicated device requiring a second flux loop and a third superconducting electrode.

\begin{figure}[tb]
	\centering
	\includegraphics[width=\columnwidth]{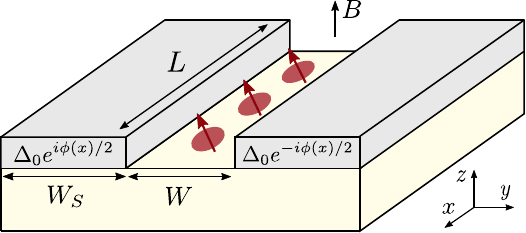}
	\caption{We consider a planar Josephson junction formed by two $s$-wave superconductors (gray) placed in contact with a spin-orbit coupled semiconductor (yellow). A perpendicular magnetic field creates a linear phase gradient along the junction, leading to an array of Josephson vortices with cores located at the positions where the superconducting phase difference is an odd multiple of $\pi$ (red dots). These Josephson vortices host low-energy bound states that can, in certain parameter regimes, be exploited to define Andreev spin qubits (red arrows).
    }
	\label{fig:setup}
\end{figure}

Here, we propose an alternative architecture for ASQs based on a planar JJ formed by two $s$-wave superconductors placed in contact with a semiconductor two-dimensional electron gas (2DEG) with strong Rashba spin-orbit coupling (RSOC), see Fig.~\ref{fig:setup}. In the presence of a weak out-of-plane magnetic field, the superconducting phase difference $\phi(x)$ varies linearly along the junction~\cite{Barone1982}, leading to the emergence of an array of Josephson vortices with cores located at the positions where $\phi(x)$ is an odd multiple of $\pi$~\cite{Cuevas2007}. In junctions based on 3D topological insulators, these Josephson vortices have been predicted to bind topologically protected Majorana zero modes~\cite{Fu2008,Potter2013,Grosfeld2011,Laubscher2024,Yue2023,Schluck2024}. Here, we show that even junctions based on topologically trivial spin-orbit coupled semiconductors can host near-zero-energy vortex states~\cite{Moehle2022,Banerjee2023} that are energetically well-separated from other low-energy states. In certain parameter regimes, these states give rise to a spinful low-energy degree of freedom that can be used to define a novel type of ASQ, which we dub the {\it vortex spin qubit} (VSQ).

Since VSQs are confined by virtue of the superconducting phase gradient, no gate electrodes are, at least in principle, required to define the `vortex dot'. Additionally, multiple vortices and, therefore, multiple VSQs can be generated in a single JJ. As such, VSQs offer prospects for a substantial reduction in device complexity and hardware overhead compared to conventional ASQ implementations. At the same time, as we show below, VSQs maintain the control advantages of conventional ASQs, allowing for supercurrent-based readout, single-qubit gates, and two-qubit gates.

\textit{Model.---}%
We consider a planar JJ formed by two $s$-wave superconductors placed on the surface of a semiconductor 2DEG with strong RSOC, see Fig.~\ref{fig:setup}. In the presence of a perpendicular magnetic field of strength $B$, the JJ can be described by the Bogoliubov-de Gennes (BdG) Hamiltonian $H=\frac{1}{2}\int d\boldsymbol{r}\,\Psi^\dagger\mathcal{H}\Psi$ with $\Psi=(\psi_\uparrow,\psi_\downarrow,\psi_\downarrow^\dagger,-\psi_\uparrow^\dagger)^T$ and
	\begin{equation}
	\mathcal{H}=\Big[\frac{\hbar^2\boldsymbol{\pi}^2}{2m}-\alpha(\pi_x\sigma_y-\pi_y\sigma_x) -\mu\Big]\tau_z+[\Delta(x,y)\tau_++\mathrm{H.c.}],\label{eq:H_cont}
	\end{equation} 
	where $\sigma_{x,y,z}$ ($\tau_{x,y,z}$) are Pauli matrices acting in spin (particle-hole) space, $\boldsymbol{\pi}=(\pi_x,\pi_y)=(-i\partial_x-\frac{e}{\hbar} A_x\tau_z,-i\partial_y)$ is the vector of momentum with $\boldsymbol{A}=(A_x,0)$ the vector potential in the Landau gauge, $m$ is the effective mass, $\mu$ is the chemical potential, $\alpha$ is the strength of the RSOC, $\tau_\pm=(\tau_x\pm i\tau_y)/2$, and $\Delta(x,y)$ is the position-dependent proximity-induced superconducting gap. We assume that the magnetic field acts only in the normal region of the JJ, such that the vector potential takes the form $A_x(y)=-BW/2$ for $y<-W/2$, $A_x(y)=By$ for $|y|<W/2$, and $A_x(y)=BW/2$ for $y>W/2$, where $W$ is the width of the junction. The superconducting gap is taken to be
	\begin{equation}
	\Delta(x,y)= 
	\Delta_0 e^{i\mathrm{sgn}(y)\phi(x)/2}\Theta(|y|-W/2),\label{eq:scgap}
	\end{equation}
	where $\Delta_0$ is real, $\phi(x)$ is the superconducting phase difference between the two superconductors, and $\Theta$ is the Heaviside step function. In the limit where the length of the junction $L$ is much smaller than the Josephson penetration length, the phase becomes a linear function of the position along the junction with a slope set by the magnetic field~\cite{Barone1982}, $\phi(x)=(2\pi\Phi/L\Phi_0)x+\phi_0$, where $\Phi$ is the magnetic flux piercing the junction, 
    $\Phi_0=h/2e$ is the flux quantum, and $\phi_0$ is a global phase difference that can be controlled, e.g., via an external flux loop.
 
 \textit{Low-energy vortex states.---}%
 We are interested in the spectrum of low-energy bound states near the center of an isolated vortex $\phi(x)\sim\pi$. The emergence of such states can be understood from a simple analytical argument starting from a translationally invariant JJ of constant phase difference $\phi(x)\equiv\pi$ in the absence of RSOC. At high transparency, such a $\pi$-JJ hosts a pair of spin-degenerate eigenstates with energy $\epsilon(k)\sim 0$~\cite{Beenakker1991} over a range of $|k|\lesssim \sqrt{2 m \mu}$ (here $k$ is the momentum along the wire, which can be viewed as a parameter in an 
 otherwise point-like $\pi$-JJ). Typically, a finite transparency at the superconductor--normal interfaces shifts this dispersion away from zero to $\epsilon(|k|\lesssim \sqrt{2 m \mu})>0$.
However, for a suitable range of $W$ and $\mu$, see the  Supplemental Material (SM)~\cite{SM} for details, the scattering from the  two interfaces can interfere, leading to a $\pi$ phase shift in the effective scattering and, ultimately, to a change in sign $\epsilon(|k|\lesssim \sqrt{2 m \mu})<0$.
The interference effect disappears at $|k|\gg \sqrt{2 m \mu}$, restoring the positive sign of the bound state energy at large $|k|$. This give rise to a pair of ``Fermi points" $\pm k_F$ with $\epsilon(\pm k_F)=0$, where the dispersion of the $\pi$-JJ states crosses zero energy.

The vortex spectrum can now be understood by considering a linear variation of the phase $\phi(x)\sim \pi$ around the center of the vortex. Working in the basis of the $\pi$-JJ states at perfect transparency, the phase variation leads to a local energy shift $\Lambda x\rho_z$, where $\rho_z$ is the Pauli-$z$ matrix acting in the space of the $\pi$-JJ states and $\Lambda$ is an effective mass. Combining this term with the interface scattering matrix element discussed in the previous paragraph---which splits the $\pi$-JJ states away from zero energy---leads to an effective 1D Hamiltonian
 \begin{equation}
 \mathcal{H}_\mathrm{eff}=\epsilon(k)\rho_x-i\Lambda\rho_z\partial_k\label{eq:H_linear}
 \end{equation}
 for each spin sector, where $\rho_x$ is the Pauli-$x$ matrix acting in the space of the $\pi$-JJ states and $x\sim i\partial_k$. By focusing on the vicinity of a Fermi point $\pm k_F$, the Jackiw-Rebbi theorem~\cite{Jackiw1976} then predicts the existence of a zero-energy bound state localized near $x\sim 0$ in real space and $k\sim \pm k_F$ in momentum space. As such, each branch in the dispersion of the $\pi$-JJ states gives rise to a zero-energy mode, yielding four zero-energy modes per Josephson vortex in total. These zero modes bear similarities with the low-energy states that have recently been discussed for vortices in JJ rings with likely low transparency~\cite{Sanjose2025}.

 When $\mu$ is sufficiently large, the hybridization between different Fermi points $\pm k_F$ is negligible. In this regime, a spin splitting between states at the same Fermi point can be generated by RSOC. Importantly, this requires both a parallel (along the junction) and a transversal (perpendicular to the junction) component of the RSOC. Indeed, unless the spin-orbit coupling breaks both mirror symmetries $M_x$ and $M_y$, one can find a time-reversal-like symmetry $\Theta_x=i\tau_z\sigma_y M_x K$ or $\Theta_y=i\sigma_y M_y K$ (here $K$ is the complex conjugation) that enforces a Kramers degeneracy and thus prevents a spin splitting. In the SM~\cite{SM}, we show that the spin splitting scales as $\delta \epsilon\propto \alpha_x\alpha_y$ to lowest order in the RSOC, where we have used $\alpha_{x,y}$ to denote the component of the RSOC along $x$ and $y$, respectively. In summary, at large $\mu$, we thus expect the low-energy BdG spectrum near an isolated Josephson vortex to consist of four pairwise degenerate near-zero-energy states at energies $\pm\delta\epsilon$.

\begin{figure}[tb]
	\centering
	\includegraphics[width=\columnwidth]{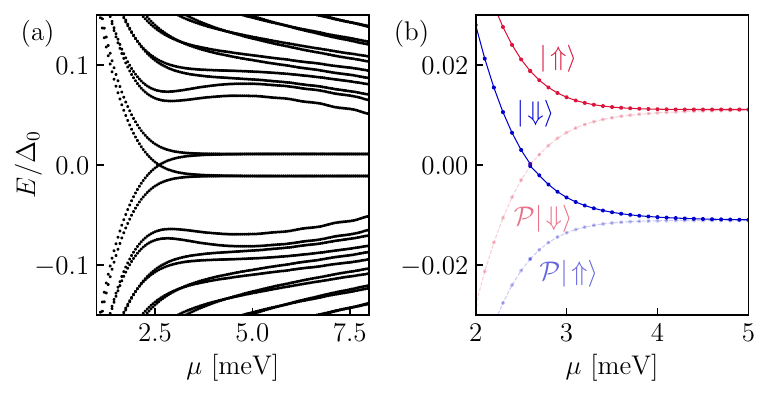}
	\caption{Numerical low-energy BdG spectrum of a planar JJ with a single Josephson vortex as a function of chemical potential $\mu$. (a) Four low-energy states, separated from the other states by an energy gap, appear for sufficiently large $\mu$. The two lowest-lying states undergo a zero-energy crossing at a critical value $\mu_c$, leading to a ground state with odd fermion parity at $\mu\gtrsim\mu_c$. (b) Expanded view of the region around $\mu_c$. Here, the two colors represent a pseudo-spin degree of freedom $\sigma\in\{\Uparrow,\Downarrow\}$. Due to particle-hole symmetry $\mathcal{P}$, only half of the energy levels correspond to physically distinct states.
    }
	\label{fig:spectrum}
\end{figure}

 To confirm the intuition outlined above, and to extract quantitative information about the spin splitting of interest, we study the full 2D JJ numerically by exact diagonalization of a discretized verion of Eq.~(\ref{eq:H_cont}), see the SM for details~\cite{SM}. Throughout this paper, we use parameters that are realistic for a Nb/InAs/Nb JJ, although our qualitative findings hold more broadly. Unless specified otherwise, we set $m=0.023m_e$ (here $m_e$ is the electron mass), $\alpha=150$~meV\r{A}~\cite{Wickramasinghe2018}, $\Delta_0=1$~meV~\cite{Telkamp2024}, $L=3~\mu$m, and $W=50$~nm. 
 In Fig.~\ref{fig:spectrum}(a), we show the low-energy BdG spectrum of a JJ threaded by a flux $\Phi=1.2\Phi_0$. The global phase was set to $\phi_0=\pi$~\footnote{This is the value of $\phi_0$ that minimizes the numerical ground state energy.}, leading to a single Josephson vortex at the center ($x=0$) of the junction. For sufficiently large $\mu$, we find that the vortex induces a large number of in-gap ($E\ll\Delta_0$) Caroli-de Gennes-Matricon (CdGM) states, with the four lowest-lying states energetically separated from the rest of the spectrum. Furthermore, the two lowest-lying states undergo a zero-energy crossing at a critical chemical potential $\mu_c$, leading to a ground state with odd fermion parity at $\mu\gtrsim\mu_c$. In this regime, the lowest-energy states are pairwise degenerate, which is consistent with our analytical arguments above. In the SM~\cite{SM}, we explicitly verify that the states at $\pm \delta\epsilon$ carry opposite spin polarizations as indicated by the pseudospin index ${\sigma\in\{\Uparrow,\Downarrow\}}$ in Fig.~\ref{fig:spectrum}(b). In particular, at the center of the vortex, the state ${|\hspace{-0.08cm}\Uparrow\rangle}$ (${|\hspace{-0.08cm}\Downarrow\rangle}$) is polarized along the $+z$ ($-z$) axis~\cite{SM}.
\textit{Qubit definition.---}%
The states $\{{|\hspace{-0.08cm}\Uparrow\rangle},{|\hspace{-0.08cm}\Downarrow\rangle}\}$ define the basis states of the VSQ. From now on we focus on the odd-parity regime $\mu\gtrsim\mu_c$, such that the state ${|\hspace{-0.08cm}\Downarrow\rangle}$ is occupied by a quasiparticle when the system is in the ground state. The excited state is generated by moving this quasiparticle to the state ${|\hspace{-0.08cm}\Uparrow\rangle}$. Note that due to the built-in particle-hole symmetry $\mathcal{P}=\sigma_y\tau_y\,K$ of the BdG Hamiltonian in Eq.~(\ref{eq:H_cont}), an occupied (unoccupied) state $|\sigma\rangle$ automatically means that its particle-hole partner $\mathcal{P}|\sigma\rangle$ is unoccupied (occupied). As such, the logical VSQ states correspond to having both blue or both red states in Fig.~\ref{fig:spectrum}(b) occupied. Configurations where one blue and one red state are occupied have a different fermion parity and therefore do not mix with the VSQ states in the absence of quasiparticle poisoning.

For an Nb/InAs/Nb JJ described by the parameters used in Fig.~\ref{fig:spectrum}, we find a qubit splitting $\epsilon_q=\delta\epsilon\sim 25~\mu$eV. This splitting scales quadratically with the RSOC strength $\alpha$ to lowest order, while it is only weakly dependent on the parent superconducting gap $\Delta_0$, which mainly determines the energy splitting between the qubit states and the next-lowest CdGM states~\cite{SM}. The spin temperature for a stable Andreev spin was previously estimated near $15$~mK~\cite{Lu2025}, such that the splittings found here are large enough for the spin to be in its quantum ground state with high probability. Therefore, initialization of the VSQ in the $|\hspace{-0.08cm}\Downarrow\rangle$ state can be achieved simply by letting the system relax to its ground state.

\textit{Readout.---}%
Readout of VSQs can be performed by adapting the cQED techniques that have been developed for conventional ASQs~\cite{Hays2018,Hays2020}. Explicitly, let us consider a system where the flux loop of the planar JJ is inductively coupled to a microwave resonator with frequency $f_r$. This modifies the phase difference across the JJ to $\phi_0+\phi_{r}$ with $\phi_r=\lambda (a+a^\dagger)$, where $\lambda$ characterizes the resonator-junction coupling strength and $a^\dagger$ is the creation operator of the resonator mode. Up to second order in $\phi_r$, the shift of the resonator frequency $\delta f_i$ due to an occupied BdG eigenstate $|i\rangle$ with energy $E_i$ is then given by~\cite{Park2020,Metzger2021}
\begin{equation}
h\delta f_i=\lambda^2\frac{\partial^2E_i}{\partial\phi_0^2}+\sum_{i\neq j}g_{ij}^2\left(\frac{2}{E_{ij}}-\frac{1}{E_{ij}-hf_r}-\frac{1}{E_{ij}+hf_r}\right),\label{eq:resonator_shift}
\end{equation}
where we have defined $E_{ij}=E_i-E_j$ and $g_{ij}=\frac{\lambda\Phi_0}{2\pi}|\langle i|J|j\rangle|$ with $J=(2e/\hbar)\,\partial \mathcal{H}/\partial\phi_0$ the current operator. The difference in resonator shift between the two qubit states is then given by $\delta f=\delta f_\Uparrow-\delta f_\Downarrow$. 

\begin{figure}[tb]
	\centering
	\includegraphics[width=\columnwidth]{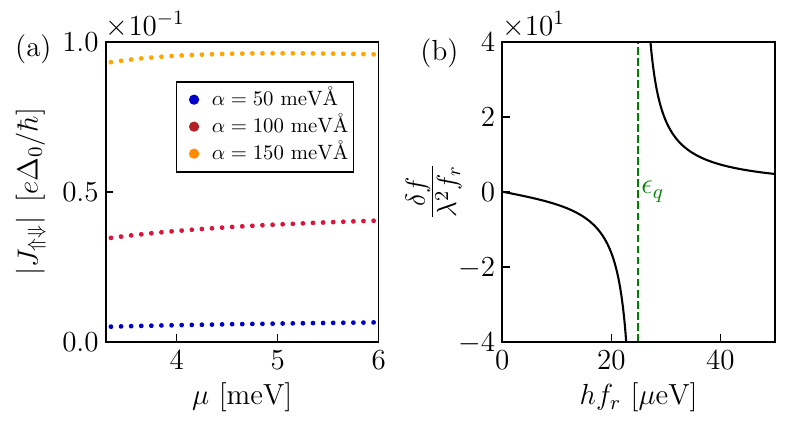}
	\caption{(a) Numerically calculated matrix element $|J_{\Uparrow\Downarrow}|$ as a function of chemical potential $\mu$ for a system with a single Josephson vortex. The different colors correspond to different RSOC strengths $\alpha$ (see inset). (b) Difference in resonator shift between the two qubit states $\delta f$ as a function of resonator frequency $f_r$ for $\alpha=150$~meV\r{A} and $\mu=5$~meV. The green line denotes the qubit splitting $\epsilon_q$. The other parameters are the same as in Fig.~\ref{fig:spectrum}.}
	\label{fig:readout}
\end{figure}

In a translationally invariant junction, the first term in Eq.~(\ref{eq:resonator_shift}) does not contribute to $\delta f$ since the energy of the vortex state does not depend on the position of the vortex and, therefore, on $\phi_0$. When the resonator frequency $f_r$ is much smaller than the energy separation between the VSQ states and the higher-energy CdGM states, $\delta f$ is largely determined by transitions between the states ${|\hspace{-0.08cm}\Uparrow\rangle}$ and ${|\hspace{-0.08cm}\Downarrow\rangle}$. In this case, we have 
\begin{equation}
h\delta f\approx  {|J_{\Uparrow\Downarrow}|^2}\left(\frac{\lambda\Phi_0}{2\pi}\right)^2\left(\frac{2}{\epsilon_q}-\frac{1}{\epsilon_q-h f_r}-\frac{1}{\epsilon_q+h f_r}\right),\label{eq:frequency_shift}
\end{equation}
where we have defined the matrix element $J_{\Uparrow\Downarrow}={\langle \Uparrow|J|\Downarrow\rangle}$. 
In Fig.~\ref{fig:readout}(a), we plot $|J_{\Uparrow\Downarrow}|$ as a function of chemical potential $\mu$ for different strengths of spin-orbit coupling $\alpha$ as obtained numerically using a discretized version of the current operator $J$~\cite{SM}. We find that $|J_{\Uparrow\Downarrow}|$ depends only weakly on chemical potential, while it increases with increasing RSOC strength. We also note that $|J_{\Uparrow\Downarrow}|$ does not depend on the global phase offset $\phi_0$ as expected in a translationally invariant junction.
In Fig.~\ref{fig:readout}(b), we show $\delta f$ as a function of $f_r$ as obtained by numerical evaluation of Eq.~(\ref{eq:frequency_shift}). Assuming for example $f_r=9$~GHz $\approx37~\mu$eV$/h$ and $\lambda=0.005$, we estimate $\delta f\approx 2$~MHz.

The readout strategy described here is particularly straightforward when there is only a single VSQ per JJ. If multiple vortices are present in a single junction, $\delta f$ will generally depend on the total multiqubit state of all VSQs. This limitation could be overcome, e.g., by slowly modulating $W$ or $\alpha$ along the junction such that each VSQ has a slightly different splitting $\epsilon_q$. The resonator frequency can then be chosen to be near-resonant with only one particular VSQ, enabling readout of individual VSQs.

\textit{Single-qubit gates.---}%
VSQ single-qubit gates can be performed by driving transitions between the qubit states either via an ac gate potential~\cite{PitaVidal2023} applied near the vortex or via an ac modulation of the flux~\cite{hays_coherent_2021}. Here, we focus on the latter, which leads to an ac variation of the global phase difference across the junction, $\phi_0(t)=\phi_0+\delta\phi_0\sin(\omega_d t)$, where $\omega_d$ is the frequency and $\delta\phi_0$ the amplitude of the drive. To first order in $\delta\phi_0\ll 1$, the system is then described by
\begin{equation}
\mathcal{H}_\mathrm{tot}(\phi_0(t))=\mathcal{H}(\phi_0)+\,\delta\phi_0\sin(\omega_d t)(\Phi_0/2\pi)\,J,
\end{equation}
where $\mathcal{H}$ is given in Eq.~(\ref{eq:H_cont}) and $J$ was defined after Eq.~(\ref{eq:resonator_shift}). For simplicity, we assume that $\omega_d$ is resonant with the qubit frequency, $\hbar\omega_d=\epsilon_q$. This leads to microwave-induced transitions between the qubit states, resulting in Rabi oscillations with Rabi frequency $hf_R=\delta\phi_0\Phi_0{|\langle \Uparrow|J|\Downarrow\rangle|}/2\pi$. The matrix element of the current operator is the same as discussed in the context of readout, see Fig.~\ref{fig:readout}(a). Assuming $\alpha=150$~meV\r{A}, $\mu=5$~meV, and $\delta\phi_0=0.01$, we estimate $f_R\approx 115$~MHz. Again, we note that the microwave drive described here will in general affect all VSQs in a particular junction simultaneously. To address individual qubits, similar techniques as discussed for readout in the previous paragraph could be applied.

\textit{Two-qubit gates.---}%
Two-qubit gates between two VSQs within the same JJ can be performed by enhancing phase fluctuations in the JJ through a combination of lowering the Josephson coupling together with driving an external ac current $I_\mathrm{ext}$ through the junction. In this case the fluctuating phase degree of freedom of the junction mediates an effective two-qubit interaction 
by shifting the energy of the qubit states by a combination of dc Lamb and ac Stark shifts. The gate operation, which is performed by adiabatically lowering the Josephson coupling and applying the ac current $I_\mathrm{ext}$ for a finite duration, is assumed to be slow compared to the qubit splitting and thus preserves the total spin along the $z$ direction, $S_z=\sigma_{z}^{(1)}+\sigma_{z}^{(2)}$ (here $\sigma_z^{(i)}$ is the Pauli-$z$ matrix acting on VSQ $i$). Furthermore, the exchange symmetry of the pair of qubits forbids any mixture between the singlet and unpolarized triplet ($S_z=0$) states since these states have opposite parities under qubit exchange.  This, combined with the degeneracy of the two $S_z=0$ states arising from the matching of the qubit frequencies, leads to a Heisenberg interaction. The effective two-qubit Hamiltonian can be written in terms of the shift of the eigenvalues of the three triplet states 
$\delta E_{T,m=0,\pm 1}$ and the singlet state $\delta E_S$ as 
\begin{align}
H_{2q}=(\mathcal{J}_I/2)\,\sigma_z^{(1)}\sigma_z^{(2)}+(\mathcal{J}_H/2)\,\bm{\sigma}^{(1)}\cdot\bm{\sigma}^{(2)},
\end{align}
where $\mathcal{J}_H=(\delta E_{T,0}-\delta E_S)/2$ and $\mathcal{J}_I=(\delta E_{T,1}+\delta E_{T,-1}-2\delta E_{T,0})/2$
 are the Heisenberg and Ising couplings respectively. The phase-fluctuation induced shifts of the eigenvalues $\delta E_{T,m=0,\pm 1}$, $\delta E_S$ are estimated in the SM~\cite{SM} for the parameters of the JJ described so far and lead to couplings of the order of tens of MHz for the dc contribution and tens to hundreds of MHz for the ac contribution. Note that the ac contributions can be enhanced by choosing the ac frequency close to the qubit splitting even when the dc shift is small. Incidentally, the phase-fluctuation induced dc Lamb shift also affects individual qubits and could be used generate a single-qubit phase gate. The dc shift can be turned off by connecting a JJ shunt across the junction to increase the Josephson coupling.

\textit{Discussion.---}%
We have proposed a novel architecture for Andreev spin qubits based on Josephson vortices in spin-orbit coupled planar Josephson junctions. We have shown that, in certain parameter regimes, the vortices can host energetically isolated spinful low-energy bound states that can be exploited to define so-called vortex spin qubits (VSQs). Single-qubit gates can be performed by flux driving, while readout can be conveniently achieved by adapting standard cQED techniques developed for conventional ASQs. We have also outlined a way in which an entangling two-qubit gate---necessary for universal quantum computation with VSQs---can be performed.

The VSQs are defined by virtue of a gradient in the superconducting phase difference across the junction, which we assume to be generated by a perpendicular magnetic field. If strong enough, such a field can destabilize the superconductor and lead to unwanted effects such as the introduction of Abrikosov vortices. This is especially relevant if, instead of Nb as in our numerical simulations, a superconductor with a smaller gap such as Al is used. As a potential alternative, the phase gradient could be generated by a dc supercurrent flowing parallel to the junction. However, it is also important to note that the actual magnetic field strength required to induce one or more VSQs in a real experimental system cannot be reliably estimated from our theoretical model since we have completely neglected the response of the superconductor to the field. In particular, to determine the flux through the junction, the junction width $W$ should be replaced by an effective width $W_\mathrm{eff}$ that takes into account a finite penetration depth and flux focusing effects. The naive estimate $B=\Phi/WL$ significantly overestimates the magnetic field needed to thread a flux $\Phi$ through the junction~\cite{Suominen2017}. Experimentally, the required field should ultimately be determined via a Fraunhofer pattern measurement.

We have neglected the effects of disorder in this work. The flatness of the vortex state energies as a function of chemical potential (see Fig.~\ref{fig:spectrum}) indicates that VSQs are likely relatively robust against long-range Coulomb-type charge disorder. On the other hand, magnetic noise due to the nuclear spin environment in InAs has been identified as a key potential source of dephasing for conventional ASQs~\cite{Hoffman2025} and will also affect the VSQs proposed here. This limitation could be overcome by switching to group IV materials, which can be isotopically purified. In this context, Ge hole-based systems constitute a particularly promising platform due to their exceptionally high  quality~\cite{Myronov2023,Lodari2022,Stehouwer2023}, large and tunable spin-orbit coupling~\cite{Winkler2003,Scappucci2020}, and compatibility with superconductivity~\cite{Hendrickx2018,Hendrickx2019,Vigneau2019,Aggarwal2021,Tosato2023}. It would be interesting to study VSQs in Ge-based planar JJs in future work.

While we have focused on a particular implementation of a two-qubit gate, where an interaction between two VSQs in the same JJ is mediated by the fluctuating phase degree of freedom, 
there are several other ways in which VSQs may interact. For example, VSQs in different JJs could be coupled via their spin-dependent supercurrents similar to conventional ASQs~\cite{Padurariu2010, PitaVidal2025,Lu2025b}. Furthermore, VSQs in the same JJ that are located sufficiently close to each other could interact via direct wave function overlap. In addition, the two-qubit interaction mediated by the fluctuating phase degree of freedom discussed above could naturally allow for multi-qubit gates acting on more than two VSQs in a given junction simultaneously. We leave a more detailed analysis of these alternative two-qubit and potential multi-qubit interactions to future work.

Finally, the quasi-1D line junctions considered here can be expanded to more complicated geometries, e.g., T junctions~\cite{Yang_2019} or junction arrays~\cite{Reinhardt2025}. This could allow one to selectively reorder vortices using flux-bias strategies similar to those recently proposed in the context of vortex Majorana zero modes~\cite{Hedge2020,Zhang2025}.

\textit{Acknowledgments.---}%
This material is based upon work supported by the Air Force Office of Scientific Research under award
number FA9550-26-1-0028. We acknowledge support by the Laboratory for Physical Sciences through the Condensed Matter Theory Center.

\bibliography{refs.bib,vf-references.bib}

\end{document}


\title{Supplemental Material: Andreev spin qubits bound to Josephson vortices in spin-orbit coupled planar Josephson junctions}

\author{Katharina Laubscher}
\affiliation{Condensed Matter Theory Center and Joint Quantum Institute, Department of Physics, University of Maryland, College Park, Maryland 20742, USA}
\author{Valla Fatemi}
\affiliation{School of Applied and Engineering Physics, Cornell University, Ithaca, NY, 14853, USA}
\author{Jay D. Sau}
\affiliation{Condensed Matter Theory Center and Joint Quantum Institute, Department of Physics, University of Maryland, College Park, Maryland 20742, USA}

\maketitle

\section{Discretized model of the junction}
\label{sec:tb}

The real-space discretized BdG Hamiltonian for the planar SNS junction in the presence of a magnetic field takes the form
\begin{align}
\bar{H}&=\frac{1}{2}\sum_{n,m}\Big\{\Big[\frac{i\alpha}{2a}\big(\boldsymbol{c}_{n+1,m}^\dagger e^{i\varphi_m\tau_z}\tau_z\sigma_y\boldsymbol{c}_{n,m}-\boldsymbol{c}_{n,m+1}^\dagger\tau_z\sigma_x\boldsymbol{c}_{n,m})-t(\boldsymbol{c}_{n+1,m}^\dagger e^{i\varphi_m\tau_z}\tau_z\boldsymbol{c}_{n,m}+\boldsymbol{c}_{n,m+1}^\dagger\tau_z\boldsymbol{c}_{n,m})\nonumber\\&\hspace{17mm}+\Delta_{n,m}\boldsymbol{c}_{n,m}^\dagger\tau_+\boldsymbol{c}_{n,m}+\mathrm{H.c.}\Big]-\mu\,\boldsymbol{c}_{n,m}^\dagger\tau_z\boldsymbol{c}_{n,m}+4t\boldsymbol{c}_{n,m}^\dagger\tau_z\boldsymbol{c}_{n,m}\Big\}\label{eq:H_tb}
\end{align}
with $\boldsymbol{c}_{n,m}=(c_{n,m,\uparrow}, c_{n,m,\downarrow}, c_{n,m,\downarrow}^\dagger, -c_{n,m,\uparrow}^\dagger)^T$ and $t=\hbar^2/2ma^2$, where $a$ is the lattice spacing. Throughout this paper, we set $a=5$~nm. Furthermore, $\Delta_{n,m}$ is the site-dependent pairing amplitude corresponding to a discretized version of the spatially dependent superconducting gap given in Eq.~(2) of the main text, and we have introduced Peierls phases arising due to the magnetic field $\varphi_m=\frac{ea}{\hbar}A_x(y)$ with $y=am$.

\begin{figure*}[tb]
	\centering
	\includegraphics[width=\columnwidth]{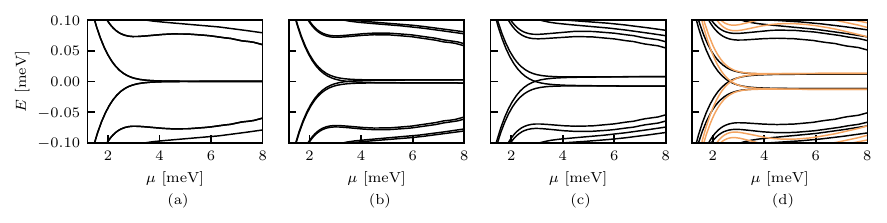}
	\caption{Numerical low-energy BdG spectrum of a JJ with a single Josephson vortex as a function of chemical potential $\mu$. (a) In the absence of spin-orbit coupling ($\alpha=0$) four zero-energy states emerge as the chemical potential reaches a critical value $\mu_\mathrm{crit}$. (b) $\alpha=50$~meV\r{A}. (c) $\alpha=100$~meV\r{A}. (d) $\alpha=150$~meV\r{A}. Panels (b-d) show that the spin-orbit coupling induces a spin splitting that scales approximately quadratically with $\alpha$. In panel (d), the black lines correspond to $\Delta_0=1$~meV, whereas the orange lines correspond to $\Delta_0=1.2$~meV. We find that the qubit splitting is only weakly affected by the parent gap, which however influences the splitting between the qubit states and the higher-energy CdGM states. The other parameters are the same as in Fig.~2 of the main text.}
	\label{fig:spectrum_sm}
\end{figure*}

In Fig.~\ref{fig:spectrum_sm}, we show example spectra obtained by numerical diagonalization of Eq.~(\ref{eq:H_tb}) for a JJ with a single Josephson vortex as a function of chemical potential. In the absence of spin-orbit coupling ($\alpha=0$), see Fig.~\ref{fig:spectrum_sm}(a), four zero-energy states emerge as the chemical potential reaches a critical value $\mu_\mathrm{crit}$ as predicted by our analytical arguments presented in the main text and detailed in Sec.~\ref{sec:analytics}. For finite spin-orbit coupling, a gap that scales approximately quadratically with $\alpha$ is opened in the spectrum of these zero-energy states, see Fig.~\ref{fig:spectrum_sm}(b)-Fig.~\ref{fig:spectrum_sm}(d). The other system parameters are the same as in the main text.

We also introduce a discretized version of the current operator, which we use to numerically calculate the matrix element $|J_{\Uparrow\Downarrow}|$ defined in the main text. At the center of the junction at $y=0$, the current density along the $x$ direction is zero, allowing us to focus on the current density operator along the $y$ direction, which can be written as
\begin{equation}
\bar{J}=-\frac{e}{\hbar}\sum_n\Big[\frac{it}{2} \boldsymbol{c}_{n,m}^\dagger\left(\boldsymbol{c}_{n,m+1}-\boldsymbol{c}_{n,m-1}\right)-\frac{\alpha}{4 a}\boldsymbol{c}_{n,m}^\dagger\sigma_x\left(\boldsymbol{c}_{n,m+1}+\boldsymbol{c}_{n,m-1}\right)\Big]+\mathrm{H.c.}
\end{equation}
The first term comes from the usual kinetic energy, while the second term is an additional contribution due to spin-orbit coupling.

\section{Spin texture of the low-energy vortex states}
\label{sec:spin}
 
\begin{figure*}[tb]
	\centering
	\includegraphics[width=\columnwidth]{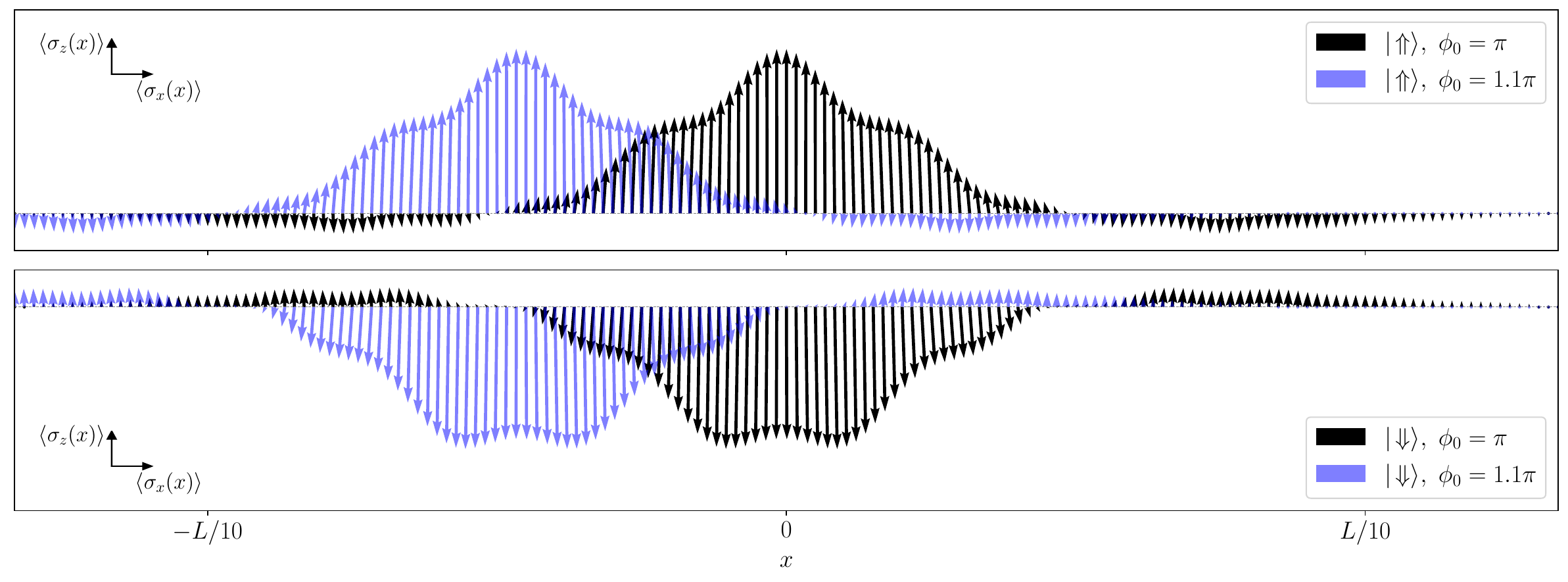}
	\caption{Numerically calculated spin polarization of the low-energy vortex states integrated over the $y$ direction and spatially resolved along the $x$ direction [see Eqs.~(\ref{eq:spinx}) and (\ref{eq:spinz})] for two different values of the global phase offset $\phi_0$. (a) State ${|\hspace{-0.08cm}\Uparrow\rangle}$. (b) State ${|\hspace{-0.08cm}\Downarrow\rangle}$. The spin polarization at the center of the vortex is fixed to $\pm z$ due to inversion symmetry. Away from the vortex center, the spin polarization rotates around the $y$ axis on the scale of the spin-orbit length $l_{so}$. As long as the vortex is located far from the junction ends, the system is effectively translationally invariant at low energies, such that the vortex states at different values of $\phi_0$ are simply related to each other by a translation along $x$. The chemical potential was set to $\mu=5$~meV. The other parameters are the same as in Fig.~2 of the main text.}
	\label{fig:spin}
\end{figure*}

In Fig.~\ref{fig:spin}, we show the spin polarization of the low-energy vortex states labeled ${|\hspace{-0.08cm}\Uparrow\rangle}$ and ${|\hspace{-0.08cm}\Downarrow\rangle}$ in Fig.~2 of the main text as a function of position $x$ along the junction for two representative values of the global phase offset $\phi_0$. Explicitly, we plot
\begin{align}
\langle\sigma_x(x)\rangle=\int dy\,\psi^\dagger_j(x,y)\sigma_x \psi_j(x,y),\label{eq:spinx}\\
\langle\sigma_z(x)\rangle=\int dy\,\psi^\dagger_j(x,y)\sigma_z \psi_j(x,y),\label{eq:spinz}
\end{align}
where $\psi_j(x,y)$ for $j\in\{\Uparrow,\Downarrow\}$ are four-component Nambu wave functions corresponding to the states ${|\hspace{-0.08cm}\Uparrow\rangle}$ and ${|\hspace{-0.08cm}\Downarrow\rangle}$. We note that the spin polarization along the $y$ direction vanishes upon integration along the $y$ axis, such that we omit it here. As such, the vortex states have a spin texture that effectively rotates around the $y$ axis as we move along the junction, consistent with the presence of a spin-orbit term $\alpha k_x\sigma_y$. Indeed, the length scale on which the spin flips once from $+z$ to $-z$ is given by $\pi l_{so}/2$, where $l_{so}=\hbar^2/m\alpha$ is the spin-orbit length. Furthermore, since the total Hamiltonian has an inversion symmetry about the center of the vortex, $\mathcal{H}(x,y)\tau_z\sigma_z=\tau_z\sigma_z\mathcal{H}(-x,-y)$, the spin polarization at the center of the vortex is fixed to the $z$ direction. As long as the vortex is located far from the junction ends, the system is effectively translationally invariant at low energies, such that the vortex states at different values of $\phi_0$ are simply related to each other by a translation along $x$.

\section{Two-qubit gate}
\label{sec:twoqubit}

Here, we show that a two-qubit interaction between two VSQs in the same JJ can be mediated by the fluctuating phase degree of freedom of the junction. The coupling between the VSQs and the fluctuating phase can be turned on by lowering the Josephson coupling of the junction, e.g., by connecting the JJ to a second, auxiliary JJ. Additionally, we consider an external ac current drive $I_\mathrm{ext}$, which allows us to further control---and resonantly enhance---the two-qubit interaction mediated by the fluctuating phase. We describe the total system consisting of the junction coupled to the bosonic phase degree of freedom by the Hamiltonian
\begin{equation}
H_\mathrm{tot}= H+\frac{\Pi^2}{2M}+K\phi_0^2+\frac{\partial{H}}{\partial\phi_0}\phi_0+\frac{\Phi_0}{2\pi}I_\mathrm{ext}(t)\phi_0,
\end{equation}
where the first term is the bare junction Hamiltonian given in Eq.~(1) of the main text. The second and third term describe the phase degree of freedom, where $\Pi$ is the conjugate momentum to $\phi_0$ with $[\phi_0,\Pi]=i$, and where $M$ and $K$ are effective parameters describing the dynamics of $\phi_0$. The fourth term describes the coupling between the junction and the phase degree of freedom, and the fifth term describes the coupling to the external current drive. We can simplify this Hamiltonian by shifting $\phi_0'= \phi_0+(\Phi_0/2\pi)I_\mathrm{ext}/2K$, which yields
\begin{equation}
H_\mathrm{tot}=H+\frac{\Pi^2}{2M}+K{\phi_0'}^2+\frac{\partial{H}}{\partial\phi_0'}\phi_0'-\frac{\partial{H}}{\partial\phi_0'}\frac{\Phi_0}{2\pi}\frac{I_\mathrm{ext}}{2K}-\left(\frac{\Phi_0}{2\pi}\right)^2\frac{I_\mathrm{ext}^2}{4K}.
\end{equation}
We first study the system in the absence of the external drive, $I_\mathrm{ext}=0$, which will be taken into account further below. We start by diagonalizing the terms describing the phase degree of freedom by introducing ladder operators $a,a^\dagger$ such that $\Pi^2/2M+K\phi_0'^2=\hbar\omega a^\dagger a$ with $\omega=\sqrt{2K/M}/\hbar$. In terms of these ladder operators, we have $\phi_0'=(1/8MK)^{1/4}(a+a^\dagger)$. Projecting the junction Hamiltonian onto the low-energy vortex states, the total Hamiltonian in the absence of the current drive can then be written as
\begin{equation}
H_{\mathrm{tot}}(I_\mathrm{ext}=0)=\frac{\epsilon_q}{2}(\sigma_z^{(1)}+\sigma_z^{(2)})+\hbar\omega a^\dagger a+J_0(\sigma_x^{(1)}+\sigma_x^{(2)})(a+a^\dagger),\label{eq:twoqubit}
\end{equation}
where $\sigma_{x,y,z}^{(i)}$ are Pauli matrices acting on VSQ $i\in\{1,2\}$, $\epsilon_q$ is the qubit splitting of the individual VSQs, and $J_0=(1/8MK)^{1/4}\,\Phi_0|J_{\Uparrow\Downarrow}|/2\pi$ is proportional to the offdiagonal matrix element of the current operator calculated in the main text. Note that here we have neglected a spin-independent diagonal matrix element of the current operator, which would only result in a global energy shift.

The Hamiltonian in Eq.~(\ref{eq:twoqubit}) is symmetric under the exchange $Q$ of VSQs 1 and 2 and conserves the total parity $P=\sigma_z^{(1)}\sigma_z^{(2)}(-1)^n$, where $n=a^\dagger a$. In the following, we consider only terms up to second order in $J_0$, which allows us to restrict the Hilbert space to $n\in\{0,1\}$.
The two $Q=-1$ sectors, which just contain the singlet states $\propto|\Uparrow\Downarrow-\Downarrow\Uparrow,\,n=0,1\rangle$, are not affected by the coupling to the fluctuating phase. The two $Q=1$ sectors are spanned by the states
\begin{align}
&Q=1,P=1:\quad|\sigma_z^{(1)}=\sigma_z^{(2)},\,n=0\rangle\oplus|\Uparrow\Downarrow+\Downarrow\Uparrow,\,n=1\rangle,\\
&Q=1,P=-1:\quad|\sigma_z^{(1)}=\sigma_z^{(2)},\,n=1\rangle\oplus|\Uparrow\Downarrow+\Downarrow\Uparrow,\,n=0\rangle,
\end{align}
and described by the Hamiltonians
\begin{equation}
H_{Q=1,P=1}=\begin{pmatrix}-\epsilon_q&J_0&0\\J_0&\hbar\omega&J_0\\0&J_0&\epsilon_q\end{pmatrix},\qquad
H_{Q=1,P=-1}=\begin{pmatrix}-\epsilon_q+\hbar\omega&J_0&0\\J_0&0&J_0\\0&J_0&\epsilon_q+\hbar\omega\end{pmatrix}
,\label{eq:Hmatrix}
\end{equation}
respectively. In the following, we assume that $\hbar\omega\gg \epsilon_q$. By diagonalizing the Hamiltonians in Eq.~(\ref{eq:Hmatrix}), we can then calculate the correction $\delta E^{DC}_\psi$ to the eigenenergies of the lowest-energy triplet states $|T,1\rangle=|\Uparrow\Uparrow,0\rangle$,  \mbox{$|T,-1\rangle=|\Downarrow\Downarrow,0\rangle$}, and \mbox{$|T,0\rangle\propto|\Uparrow\Downarrow+\Downarrow\Uparrow,0\rangle$,} respectively. The lowest-energy singlet state \mbox{$|S\rangle\propto|\Uparrow\Downarrow-\Downarrow\Uparrow,0\rangle$}, on the other hand, remains at zero energy. We note that in the above we have focused solely on the energies, while we have not discussed any of the changes to the eigenstates that occur while the coupling is turned on. This is  because of the adiabaticity of the gate protocol where the perturbation is turned off at the end of the gate, which in turn ensures that
the eigenstates necessarily come back to themselves at the end of the process. The only effect of the gate on the states is a phase factor $e^{-i\delta E_s t}$, which  is determined by the calculated energy eigenvalue $\delta E_s$ ($s$ is a label for one of the 4 states).  The only mixing that could in principle occur adiabatically would be between the $|T,0\rangle$ and $|S\rangle$ state, which is however forbidden by the parity eigenvalues $P$ of the two states being $\pm 1$ respectively.
The unitary time-evolution associated with the adiabatic protocol is then diagonal in the basis of states $\ket{s}=\{\ket{T,m=0,\pm 1},\ket{S}\}$ with the eigenvalues $e^{-i\delta E_s t}$. The unitary time-evolution operator can now be written in terms of an effective two-qubit Hamiltonian 
\begin{align}
    e^{-i H_{2q}t}=\sum_s e^{-i\delta E_s t}\ket{s}\bra{s},
\end{align}
where 
\begin{align}
    &H_{2q}=\sum_s \delta E_s \ket{s}\bra{s}=(\mathcal{J}_I/2)\sigma_z^{(1)}\sigma_z^{(2)}+(\mathcal{J}_H/2)\bm\sigma^{(1)}\bm\sigma^{(2)},\label{eq:SM2Q}
\end{align}
and $\mathcal{J}_I$ and $\mathcal{J}_H$ are Ising and Heisenberg couplings. 
Focusing on the case of the dc perturbation $J_0$ for now, the strength of the dc two-qubit interaction that results from adiabatically turning on the coupling to the fluctuating phase can then be extracted as
\begin{align}
\mathcal{J}_I^{DC}&=(\delta E_{T,+1}^{DC}+\delta E_{T,-1}^{DC}-2\delta E_{T,0}^{DC})/2,\\
\mathcal{J}_H^{DC}&=(\delta E_{T,0}^{DC}-\delta E_{S}^{DC})/2.
\end{align}
To understand this correspondence with eigenvalues, we calculate the eigenvalues of the 4 states $\ket{s}$ using the Hamiltonian in  Eq.~(\ref{eq:SM2Q}) to be 
\begin{align}
    \{\mathcal{J}_H/2-\mathcal{J}_I/2,\mathcal{J}_H/2+\mathcal{J}_I/2,\mathcal{J}_H/2+\mathcal{J}_I/2,-3\mathcal{J}_H/2-\mathcal{J}_I/2 \}=\{\delta E_{T,0},\delta E_{T,1},\delta E_{T,-1},\delta E_S\}.
\end{align}

Next, we take the external current drive into account. In the presence of an external ac current $I_\mathrm{ext}(t)=I_\mathrm{ext}^0\cos{\Omega t}$, each eigenstate additionally experiences an ac Stark shift~\cite{delone1999ac} given by
\begin{equation}
\delta E_\psi^{AC}\approx -\frac{\Phi_0}{2\pi}\frac{I_\mathrm{ext}^0}{2K}\sum_{\psi\neq\psi'}\frac{|\langle\psi'|J_1|\psi\rangle|^2(E_{\psi'}-E_\psi)}{(E_{\psi'}-E_\psi)^2-\hbar\Omega^2},
\end{equation}
where $|\psi\rangle$ denotes an eigenstate obtained from Eq.~(\ref{eq:Hmatrix}) with energy $E_\psi$. Here, the operator
\begin{equation}
J_1=\frac{\Phi_0|J_{\Uparrow\Downarrow}|}{2\pi}\begin{pmatrix}0 &1&0\\1&0&1\\0&1&0\end{pmatrix}
\end{equation}
couples states with opposite $P$ quantum numbers, i.e., states belonging to different blocks in Eq.~(\ref{eq:Hmatrix}). The ac Stark shift results in additional energy shifts to the low-energy triplet states, whereas the singlet state again remains at zero energy. The strength of the two-qubit interaction generated by adiabatically turning on the ac current drive is then given by
\begin{align}
\mathcal{J}_I^{AC}&=(\delta E_{T,1}^{AC}+\delta E_{T,-1}^{AC}-2\delta E_{T,0}^{AC})/2,\\
\mathcal{J}_H^{AC}&=(\delta E_{T,0}^{AC}-\delta E_{S}^{AC})/2.
\end{align}
To estimate the strength of the two-qubit interaction, we use the qubit splitting $\epsilon_q\sim25~\mu$eV and current matrix element $\Phi_0|J_{\Uparrow\Downarrow}|/2\pi\sim 50~\mu$eV from the main text. We further estimate $K\sim k_F W\Delta_0\sim 3$~meV when the coupling to the phase degree of freedom is turned on, which leads to $J_0\sim 5~\mu$eV. Setting $\hbar\omega=5\epsilon_q$ then leads to DC coupling strengths of  $\mathcal{J}_I^{DC}/h\sim 50$~MHz and $\mathcal{J}_H^{DC}/h\sim -50$~MHz. In order to estimate the ac contribution, we assume a near-resonant driving frequency $\hbar\Omega=0.8\epsilon_q$ and a driving current amplitude of $I_\mathrm{ext}^0\sim 0.05I_c$ with $I_c=(2\pi/\Phi_0)K$. This yields $\mathcal{J}_I^{AC}/h\sim 215$~MHz and $\mathcal{J}_H^{AC}/h\sim -70$~MHz. These couplings can be further controlled by tuning $\hbar\Omega$ towards or away from the resonance, or by increasing or decreasing $I_\mathrm{ext}^0$.

\section{Vortex spectrum and splitting}
\label{sec:analytics}

The purpose of this section is to consider a simplified model where we can develop an analytic understanding of the emergence of zero-energy modes at Josephson vortices [see Fig.~\ref{fig:spectrum_sm})(a)] and the splitting of these zero modes due to spin-orbit coupling [see Figs.~\ref{fig:spectrum_sm}(b)-\ref{fig:spectrum_sm}(d)]. For this purpose, let us consider the spectrum of a Josephson vortex in a $\pi$ junction in a spin-orbit coupled semiconductor. The spin-orbit coupling is chosen to be $H_{SO}=\alpha k_y \sigma_x-\alpha' k_x\sigma_y$ with $\alpha'\ll\alpha$, allowing us to treat $\alpha'$ perturbatively.

For convenience, let us first rotate the spin matrices from $\sigma_x\rightarrow\sigma_y$ and $\sigma_y\rightarrow -\sigma_x$ so that the Hamiltonian of the $\pi$ junction  with a vortex is
\begin{equation}
H_0=[k^2-\mu(y)-\partial_y^2-i\alpha\partial_y\sigma_y]\tau_z+\Delta(y)[\cos{\phi(x)}\tau_x+\sin{\phi(x)}\tau_y],
\end{equation}
where $\Delta(y)= \Theta(|y|-W/2)\mathrm{sign}(y)\Delta_0$ and $\phi(x)=\Lambda x$ (where $\Lambda=\Delta_0 (2\pi \Phi/L\Phi_0)$ in terms of variables defined in the main text).
Here $\mu(y)$ is a position dependent electro-chemical potential that will later be chosen to include barriers at the interface and $W$ is the width of the junction.

For $\alpha'=0$, the SO coupling can be removed by a unitary rotation $U=\exp[i\alpha\sigma_y y/2]$. Explicitly, since $U^\dagger\partial_y U=\partial_y+i\alpha \sigma_y/2$, we have
\begin{equation}
U^\dagger H_1 U=U^\dagger[(k^2-\mu(y)-\alpha^2/4-\partial_y^2)\tau_z+\Delta(y)\{\cos{\phi(x)}\tau_x+\sin{\phi(x)}\tau_y\}]U=H_0.
\end{equation}
The Hamiltonian $H_1$ has a symmetry $M_x K$ where $M_x$ is the $x\rightarrow -x$ mirror symmetry and $K$ is complex conjugation.

Before proceeding further, let us understand the conditions where the energy eigenvalue $\epsilon(k)$ of the $\pi$ junction without the vortex (i.e. $\Phi=0$) crosses zero-energy, which is a key requirement. Firstly, if $\mu(y)$ is constant and $\mu\gg \Delta$ (i.e. Andreev limit), defining $k_{F,y}^2=\mu+\alpha^2/4-k^2$, we can further transform 
\begin{equation}
\tilde{H}_1\equiv e^{i k_{F,y}y} H_1 e^{-i k_{F,y}y}=-i k_{F,y}\partial_y\tau_z+\Delta(y)\tau_x.
\end{equation}
Since $\Delta(y)$ has a change of sign, the Jackiw-Rebbi theorem implies there are a pair of zero energy states $\psi_k(y)\sim e^{\pm i k_{F,y}y}e^{-k_{F,y}^{-1}\int_0^y d y_1 \Delta(y_1)}(1,\pm i)^T$ (where the two components refer to particle and hole). These states are related by complex conjugation, so hence forward will be referred to as $\psi_k(y)$ and $\psi_k^*(y)$. The $\pi$ junction states are moved away from $\epsilon(k)=0$ by barriers $g\delta(y\pm W/2)$ included in $\mu(y)$. This leads to a scattering matrix element between states near $\pm k_{F,y}$ with amplitude $\epsilon(k)=2 g \cos{k_{F,y}W}$ between $\psi_k(y)$ and $\psi_k^*(y)$, where $k_{F,y}=\sqrt{\mu+\alpha^2/4-k^2}$ is implicitly a function of $k$ that vanishes and changes sign at $k_{F,y}W=\pi/2$. This demonstrates the origin of near flat band states with Fermi points at finite width $\pi$ Josephson junctions.

Next let us add the $x$ variation of the pairing phase, away from $\pi$. For small $\Phi$, this variation is slow with $x$ and locally shifts the $\pi-$JJ states $\psi_k(y)$ and $\psi_k^*(y)$ away from $0$ (even for $\epsilon(k)=0$) by $\sim \Delta_0 \phi(x)=\Lambda x$. Working in the basis where the perfect transparent (i.e.  $\epsilon(k)=0$) $\pi-$JJ states are eigenstates of the Pauli matrix $\rho_z$ and $\epsilon(k)$ scatters between these states the effective Hamiltonian including both of these terms can be written as a momentum-space Jackiw-Rebbi Hamiltonian
\begin{equation}
H_2=\epsilon(k)\rho_x-i\Lambda\rho_z\partial_k,
\end{equation}
where we have written the position operator in the momentum space as  $x=-i\partial_k$. The Hamiltonian $H_1$ with a vortex had a symmetry $M_x K$ where $M_x$ is the $x\rightarrow -x$ mirror symmetry. This symmetry preserves $k$ but exchanges $\psi_k(y)$ and $\psi_k^*(y)$. This symmetry in the new basis is $\rho_x K$ for the Hamiltonian $H_2$.
When $\epsilon(k)$ changes sign, there is a zero-mode solution formally given by the wave-function $\Psi_k=\exp[\Lambda^{-1}\rho_y\int_0^k dk_1 \epsilon(k_1) ]\Psi_0$. $\Psi_0$ has to be an eigenstate of $\rho_y$ with eigenvalue $-1$.
Expanding $\Psi_0$ in terms of the basis $\psi_k(y)$ and $\psi_k^*(y)$ yields 
\begin{align}
\Psi_k(y)=[e^{i\pi/4}\psi_k(y)+e^{-i\pi/4}\psi_k^*(y)]\exp[-\Lambda^{-1}\int_0^k dk_1 \epsilon(k_1) ],
\end{align}
which makes it explicit that $\Psi_k^*(y)=\Psi_k(y)$ as expected from the $M_x K$ symmetry. The Hamiltonian $H_1$ has a further $M_x M_y \tau_z$ symmetry where $M_y$ is the $y$ mirror sending $y\rightarrow -y$. Since $M_x$ acts trivially on $k$ this ensures $\Psi_k(-y)=\tau_z\Psi_k^*(y)$.

As a final step we now consider the splitting of the zero-mode from the $x$-SO coupling $\alpha'$. Accounting for the transformation $U$ required to transform the state $\Psi_k(y)$ to the eigenstate $U\Psi_k(y)$ of $H_0$, the relevant matrix element is written as
\begin{equation}
(\alpha'/2)\int dk dy [ U\sigma_x\Psi_k(y)]^\dagger k\sigma_x\tau_z U\Psi_k(y),
\end{equation}
where the $\sigma_x$ in the left state creates a $\sigma_y=-1$ eigenstate from the $\sigma_y=+1$ state that $\Psi_k$ is assumed to be. In the absence of $y$-SO coupling, i.e. $\alpha=0$ (and $U=\bm 1$), the matrix element vanishes because $\Psi^\dagger_k(y) \tau_x\Psi_k(y)=\Psi^T_k(-y) \tau_z\tau_x\tau_z\Psi^*_k(-y)=-\Psi^\dagger_k(-y) \tau_x\Psi_k(-y)$. However, one can combine the spin parts of the matrix elements to $\sigma_x U^\dagger \sigma_x U=U^2\simeq \mathbf{1} -\alpha \sigma_y y$. 
Therefore, to linear order, the matrix element, which is also the energy splitting, is
\begin{equation}
\delta \epsilon=\alpha\alpha'\int dk dy \Psi^T_k(y) yk\tau_x \Psi_k(y)\exp[\Lambda^{-1}\int_0^k dk_1 \epsilon(k_1) ],
\end{equation}
where the extra $y$ prevents the integral from vanishing under the $M_y$ symmetry.
The energy splitting $\delta \epsilon$ is the matrix element between the different $\sigma_y=\pm 1$ states. This splits the zero-modes in a symmetric way to energies $\pm \delta \epsilon$. A similar phenomenon occurs at each of the Fermi points $\pm k_F$, which are decoupled leading to the structure at large $\mu$ in Figs.~\ref{fig:spectrum_sm}(b)-\ref{fig:spectrum_sm}(d). At small $\mu$ the different Fermi points hybridize as well, leading to other crossings.

\bibliography{refs.bib,vf-references.bib}